Suggestion of the DLV dimensionless number system to represent the scaled behavior of structures under impact loads

Shuai WANG [a,b], Fei XU [a,b,*], Zhen DAI [a,b]

[a] School of Aeronautics, Northwestern Polytechnical University, Xi'an 710072, Shannxi, P.R. China.

[b] Institute for Computational Mechanics and Its Applications, Northwestern Polytechnical University, Xi'an 710072, Shannxi, P.R. China.

[*] Correspondence information: Fei XU, E-mail address: xufei@nwpu.edu.cn

Suggestion of the DLV dimensionless number system to represent the scaled behavior of structures under impact loads

Shuai WANG [a,b], Fei XU [a,b,*], Zhen DAI [a,b]

[a] School of Aeronautics, Northwestern Polytechnical University, Xi'an 710072, Shannxi, P.R. China.

[b] Institute for Computational Mechanics and Its Applications, Northwestern Polytechnical University, Xi'an 710072, Shannxi, P.R. China.

[*] Corresponding author. E-mail address: xufei@nwpu.edu.cn

Highlights

(1) The DLV dimensionless number system to represent structural impact is proposed.

(2) Two well-known numbers, the damage number and the response number, are naturally included in DLV dimensionless numbers.

(3) The property of directly matching the dimensionless expression of the response equations is verified through simple equation analysis of four impact models.

(4) The ability of addressing non-scalability as well as the VSG system is confirmed.

**ABSTRACT** : A group of dimensionless numbers, termed DLV (Density-Length-Velocity) system, is put forward to represent the scaled behavior of structures under impact loads. It is obtained by means of the Buckingham Π theorem with an alternative basis. The distinct features of this group of dimensionless numbers are that it relates physical quantities of the impacted structure with essential basis of the Density, the Length and the Velocity, and thus it can represent the scaled influence of material property, geometry characteristic and velocity on the behavior of structures. The newly 15 proposed dimensionless numbers reflect three advantages. (1) The intuitively clear physical significance of these dimensionless numbers, such as the ratios of force intensity, force, moment of inertia to the corresponding dynamic quantities, the Johnson's damage number $D_n$ and Zhao's response number $R_n$ etc. are naturally included. (2) The property of directly matching the dimensionless expression of response equations of dynamic problems with these dimensionless numbers through simple equation analysis; (3) The ability of addressing non-scaling problems for different materials and strain-rate-sensitive as well as the VSG (initial impact Velocity-dynamic flow Stress-impact mass G) system. Four classical impact models are used to verify the directly matching property and the non-scaling addressing ability of the DLV system by equation analysis. The results show that the proposed dimensionless number system is simple, clear and efficient, and we suggest using it to represent the scaled behavior of structures under impact loads.

# 1. Introduction

In order to represent the scaled behavior of structures under impact loads, Jones [1] systematically summarized previous works and attempted to describe the dynamic plastic behavior of structures through 22 dimensionless numbers based on classic MLT (Mass–Length–Time) dimensional analysis. The structural similarity indicated the predictability that behavior of the prototype could be speculated through the scaling law, which usually linked same physical variables between the prototype and the scaled model by different scaling factors. For example, the geometric scaling factor was given as

$$\beta = \frac{L_m}{L_p}, \tag{1}$$

where $L$ was the characteristic length of structure, the subscript $m$ and $p$ represented the scaled model and the prototype, respectively. When these 22 dimensionless numbers were used to relate the scaled model to the prototype, the main scaling factors of physical variable presented by this single geometric factor were listed in Table 1.

Table 1　The main scaling factors of structural impact in MLT system.

| Variable | Scaling factor | Variable | Scaling factor |
| --- | --- | --- | --- |
| Length, $L$ | $\beta$ | Strain, $\sigma$ | $\beta_\sigma = 1$ |
| Mass, $M$ | $\beta_M = \beta^3$ | Stress, $\varepsilon$ | $\beta_\varepsilon = 1$ |
| Time, $t$ | $\beta_t = \beta$ | Strain rate, $\dot{\varepsilon}$ | $\beta_{\dot{\varepsilon}} = 1/\beta$ |
| Velocity, $V$ | $\beta_V = 1$ | Acceleration, $A$ | $\beta_A = 1/\beta$ |

| Displacement, $\delta$ | $\beta_\delta = \beta$ | Energy, $E$ | $\beta_E = \beta^3$ |

However, when effects of the strain rate, the gravity and the fracture of structure were taken into account, the above scaling factors would become invalid [1]. And a hidden assumption of this group of scaling factors lay in using same material between the scaled model and the prototype, thus the MLT system could not deal with the problem of different materials. The above two aspects of the classic MLT system limits the application to describe the scaled behavior of structures under impact loads.

Recently, Oshiro and Alves [2] proposed the VSG ($V$ represented initial impact velocity $V_0$, $S$ represented dynamic flow stress $\sigma_d$ and $G$ represented impact mass) dimensionless number system to represent the scaled behavior of the impacted structures, in which the dimensionless number of physical quantities based on $V_0 - \sigma_d - G$ dimensional analysis. The most prominent feature of this system is that it has the strong ability to address the non-scaling problems arising from strain-rate-sensitive, which has been verified by Ref.[2-5]. Further study of Mazzariol et al. [6] added a new dimensionless number of structural mass to address the non-scaling problem of different materials between the prototype and the scaled model. In a more complete VSG system [7-8], 8 dimensionless numbers were given to express the acceleration $A$, the time $t$, the displacement $\delta$, the strain rate $\dot{\varepsilon}$, the stress $\sigma$, the structural mass $M'$, the force $F$ and the energy $E$ as follows:

$$\frac{A^3 G}{V_0^4 \sigma_d}, \frac{t^3 \sigma_d V_0}{G}, \frac{\delta^3 \sigma_d}{G V_0^2}, \dot{\varepsilon}\left(\frac{G}{\sigma_d V_0}\right)^{1/3}, \frac{\sigma}{\sigma_d}, \frac{M'}{G}, \frac{F^3}{V_0^4 \sigma_d G^2}, \frac{E}{G V_0^2}. \qquad (2)$$

Instead of the classic MLT system, the scaling factors of main physical variables [4, 6-8] for the VSG system were listed in Table 2. It could be seen that, except for the geometric scaling factor $\beta$, two more factors $\beta_V$ and $\beta_\rho$ that containing the influence of velocity and different materials were included.

Table 2 The main scaling factors of structural impact in VSG system.

| Variable | Scaling factor | Variable | Scaling factor |
| --- | --- | --- | --- |
| Length, $L$ | $\beta = L_m/L_p$ | Displacement, $\delta$ | $\beta_\delta = \beta$ |
| Density, $\rho$ | $\beta_\rho = \rho_m/\rho_p$ | Strain, $\sigma$ | $\beta_\sigma = \beta_\rho \beta_V^2$ |
| Velocity, $V$ | $\beta_V = V_m/V_p$ | Stress, $\varepsilon$ | $\beta_\varepsilon = 1$ |
| Mass, $M$ | $\beta_M = \beta_\rho \beta^3$ | Strain rate, $\dot{\varepsilon}$ | $\beta_{\dot{\varepsilon}} = \beta_V/\beta$ |
| Time, $t$ | $\beta_t = \beta/\beta_V$ | Force, $F$ | $\beta_F = \beta_\rho \beta^2 \beta_V^2$ |
| Acceleration, $A$ | $\beta_A = \beta_V^2/\beta$ | Energy, $E$ | $\beta_E = \beta_\rho \beta^3 \beta_V^2$ |

Nonetheless, several main obvious defects still existed in the VSG system: (1) Complex expression form and less physical meaning of most dimensionless numbers. (2) Difficulty to represent various impact loading since the impact mass $G$ was chosen as one base. (3) Denaturalization that six dimensionless numbers were expressed by the dynamic flow stress $\sigma_d$ of a material property. (4) Lack of basic physical quantities that describe structure response of impact problems such as the density, the geometrical characteristic (e.g. length, width and thickness), the strain, the angle, the angular velocity, the angular acceleration and the bending moment, etc.

In this paper, the main objective is to propose a new group of dimensionless number system to overcome main defects of the previous MLT and the VSG dimensionless systems.

In what follows, Section 2 introduces our newly proposed dimensionless system including the derivation of these dimensionless numbers and the intuitive interpretation of their physical significance. Section 3 presents the scaling factors obtained by these dimensionless numbers. Section 4 verifies the features of new system through four impact models. Finally, Section 5 summaries this work.

**2. The DLV dimensionless numbers**

In order to describe the impact behavior of structure more systematically and reasonably, we now use the Buckingham $\Pi$ theorem to rederive the dimensionless system again.

The Buckingham $\Pi$ theorem [9-10] postulates that if a system containing $n$ numbers of variables $X_i$ is expressed as a function,

$$\Phi(X_1, X_2, \ldots, X_n) = 0, \tag{3}$$

in which only $k(k < n)$ numbers of variables are independent, the function can be reduced to a relationship about $n - k$ dimensionless numbers $Y_1, \ldots, Y_{n-k}$,

$$\Phi(Y_1, Y_2, \ldots, Y_{n-k}) = 0, \tag{4}$$

where each $Y_i$ is constructed from $X_1, \ldots, X_k$ by a specified form,

$$Y_i = X_1^{a_1} X_2^{a_2} \ldots X_k^{a_k}, \tag{5}$$

with the exponents $a_1, \ldots, a_k$ being the rational numbers.

Ignoring elastic effect, gravity effect, thermal effect and fracture failure, the dynamic plastic behavior of rigid-plastic materials including strain-hardening effects and strain-rate-sensitive is supposed to be mainly controlled by the following 18 interest physical variables which are density $\rho$, characteristic length $L$, velocity $V$, stress $\sigma$ (in this paper, $\sigma$ mainly presents dynamic flow stress $\sigma_d$), force $F$, bending moment $\Psi$, time $t$, strain rate $\dot{\varepsilon}$, acceleration $A$, angular velocity $\dot{\theta}$, angular acceleration $\ddot{\theta}$, energy $E$, impulse $I$, mass $M$ (e.g. structural mass $M'$ and impact mass $G$), geometrical characteristic $H'$ (e.g. thickness and width), displacement $\delta$, strain $\varepsilon$ and angle $\theta$.

According to the Buckingham $\Pi$ theorem, when the essential variables of characteristic density $\rho$, characteristic length $L$ and characteristic velocity $V$ are chosen as the base, the relationships of these 18 physical quantities can be reduced to 15 dimensionless numbers as follow,

$$\underbrace{\left[\frac{\rho V^2}{\sigma_d}\right]}_{\Pi_1}, \underbrace{\left[\frac{\rho L^2 V^2}{F}\right]}_{\Pi_2}, \underbrace{\left[\frac{\rho L^3 V^2}{\Psi}\right]}_{\Pi_3}, \underbrace{\left[\frac{tV}{L}\right]}_{\Pi_4}, \underbrace{\left[\frac{\dot{\varepsilon}L}{V}\right]}_{\Pi_5}, \underbrace{\left[\frac{AL}{V^2}\right]}_{\Pi_6}, \underbrace{\left[\frac{\dot{\theta}L}{V}\right]}_{\Pi_7},$$
$$\underbrace{\left[\frac{\ddot{\theta}L^2}{V^2}\right]}_{\Pi_8}, \underbrace{\left[\frac{E}{\rho L^3 V^2}\right]}_{\Pi_9}, \underbrace{\left[\frac{I}{\rho L^3 V}\right]}_{\Pi_{10}}, \underbrace{\left[\frac{M}{\rho L^3}\right]}_{\Pi_{11}}, \underbrace{\left[\frac{H'}{L}\right]}_{\Pi_{12}}, \underbrace{\left[\frac{\delta}{L}\right]}_{\Pi_{13}}, \underbrace{[\varepsilon]}_{\Pi_{14}}, \underbrace{[\theta]}_{\Pi_{15}}.$$
(6)

In the following context, we would explain the meaning of each number intuitively.

The number $\Pi_1$ can be interpreted as the ratio of inertia force intensity $\rho V^2$ to the resistance ability $\sigma_d$ of a material, which is well known as the damage number $D_n$ proposed by Johnson [11] and used to measure the order of strain imposed in various impact regions of a structure.

The number $\Pi_2$ can be interpreted as the ratio of inertia force $\rho L^2 V^2$ to structural dynamic force $F$.

The number $\Pi_3$ can be interpreted as the ratio of inertia moment $\rho L^3 V^2$ to structural dynamic bending moment $\Psi$.

It should be noted that $\Pi_2$ and $\Pi_3$ can be expressed as the forms of response number $R_n$ proposed by Zhao, which is widely used to measure the response of simple impacted structures [12-16]. For example, considering the influence of the fully plastic axial membrane force $F = \sigma_d BH$ and the fully plastic bending moment $\Psi = \sigma_d BH^2/4$ on the rectangular beam with density $\rho$, thickness $H$, width $B$ and initial impulsive velocity $\bar{V}_0$, the number $\Pi_2$ and $\Pi_3$ can be rewritten as

$$\Pi_2 = \frac{\rho L^2 \bar{V}_0^2}{\sigma_d BH} = \frac{\rho \bar{V}_0^2}{\sigma_d} \frac{L}{H} \frac{L}{B} \tag{7}$$

and

$$\Pi_3 = \frac{\rho L^3 \bar{V}_0^2}{\sigma_d BH^2/4} = 4 \frac{\rho \bar{V}_0^2}{\sigma_d} \left(\frac{L}{H}\right)^2 \frac{L}{B}. \tag{8}$$

It can be seen that $\Pi_2$ is the product of the form $(\rho \bar{V}_0^2/\sigma_d) \cdot (L/H)$ and the aspect ratio $L/B$; while $\Pi_3$ is the product of the form $(\rho \bar{V}_0^2/\sigma_d) \cdot (L/H)^2$ and the aspect ratio $L/B$. The expressions of $(\rho \bar{V}_0^2/\sigma_d) \cdot (L/H)$ and $(\rho \bar{V}_0^2/\sigma_d) \cdot (L/H)^2$ are two important forms of the response number $R_n$ expressed by the multiplicator of the damage number $D_n$.

The number $\Pi_4$ to $\Pi_8$ can be interpreted as the ratio of structural five physical quantities, time $t$, strain rate $\dot{\varepsilon}$, acceleration $A$, angular velocity $\dot{\theta}$ and angular acceleration $\ddot{\theta}$ to five characteristic quantities, $L/V$, $V/L$, $V^2/L$, $V/L$ and $V^2/L^2$, respectively. One obvious feature is that these physical quantities relate only to two essential variables $V$ and $L$.

The number $\Pi_9$ and $\Pi_{10}$ can be interpreted as the ratio of structural energy $E$ and impulse $I$ to characteristic quantities $\rho L^3 V^2$ and $\rho L^3 V$, respectively.

The number $\Pi_{11}$ can be interpreted as the ratio of mass $M$ (both structural mass $M'$ and impact mass $G$) to structural characteristic mass $\rho L^3$, which means the mass of each objects in an impact problem can be scaled.

The number $\Pi_{12}$ can be interpret as the ratio of a various of geometrical size $H'$ to the characteristic length $L$, which indicates that geometric dimensions at different direction can be scaled in the same ratio, so that the scaled model maintain the similarity of geometrical configuration with the prototype.

The number $\Pi_{13}$ can be interpreted as the ratio of displacement $\delta$ to the characteristic length $L$, which means that dimensionless deformation of the scaled model and those of the prototype should remain unchanged.

The number $\Pi_{14}$ and $\Pi_{15}$ can be interpreted as an invariance property of strain $\varepsilon$ and generalized strain of angle $\theta$ for the deformed similarity. It should be noted that these two numbers should be dimensionless in nature, which in any case of scaling must be guaranteed to be constant.

From the above explanation, the advantages of DLV system are: (1) Two important numbers, Johnson's damage number $D_n$ and Zhao's response number $R_n$, are included in the 15 dimensionless numbers, which has the ability to measure the damage and the response of impacted structure. (2) Since the density is used as one basis in DLV system instead of the impact mass in VSG system, the effect of different density material on the scaling behavior of the structure can be directly expressed. At the same time, the various impact loading (concentrated mass initial velocity $V_0$ or impulsive velocity $\bar{V}_0$) would be available in DLV system. (3) These dimensionless numbers are all constructed from elementary definition of physical quantities, e.g. $\Pi_5$ in DLV system is constructed by strain rate $\dot{\varepsilon}$ with simple form of $V/L$ rather than $\dot{\varepsilon}$ with complex form of $(\sigma_d V_0/G)^{1/3}$ in VSG system. (4) 18 basic physical quantities and 15 dimensionless numbers are included in DLV system more than 11 basic physical quantities and 8 dimensionless numbers in VSG system.

## 3. Scaling factors

Since the newly proposed DLV dimensionless system based on the density, the length and the velocity, three scaling factors of the density scaling factor $\beta_\rho$, the geometric scaling factor $\beta$ and the velocity scaling factor $\beta_V$ are immediately and directly used to express the factors of other physical quantities.

For perfect structural similarity, all dimensionless numbers of the scaled model must be equal to those of the prototype, which lead

$$\frac{(\Pi_1)_m}{(\Pi_1)_p} = \frac{\beta_\rho \beta_V^2}{\beta_{\sigma_d}} = 1 \rightarrow \beta_{\sigma_d} = \beta_\rho \beta_V^2, \tag{9}$$

$$\frac{(\Pi_2)_m}{(\Pi_2)_p} = \frac{\beta_\rho \beta^2 \beta_V^2}{\beta_F} = 1 \rightarrow \beta_F = \beta_\rho \beta^2 \beta_V^2, \tag{10}$$

$$\frac{(\Pi_3)_m}{(\Pi_3)_p} = \frac{\beta_\rho \beta^3 \beta_V^2}{\beta_\Psi} = 1 \rightarrow \beta_\Psi = \beta_\rho \beta^3 \beta_V^2, \tag{11}$$

$$\frac{(\Pi_4)_m}{(\Pi_4)_p} = \frac{\beta_t \beta_V}{\beta} = 1 \rightarrow \beta_t = \beta / \beta_V, \tag{12}$$

$$\frac{(\Pi_5)_m}{(\Pi_5)_p} = \frac{\beta_{\dot{\varepsilon}} \beta}{\beta_V} = 1 \rightarrow \beta_{\dot{\varepsilon}} = \beta_V / \beta, \tag{13}$$

$$\frac{(\Pi_6)_m}{(\Pi_6)_p} = \frac{\beta_A \beta}{\beta_V^2} = 1 \rightarrow \beta_A = \beta_V^2 / \beta, \tag{14}$$

$$\frac{(\Pi_7)_m}{(\Pi_7)_p} = \frac{\beta_{\dot{\theta}}\beta}{\beta_V} = 1 \to \beta_{\dot{\theta}} = \beta_V / \beta, \tag{15}$$

$$\frac{(\Pi_8)_m}{(\Pi_8)_p} = \frac{\beta_{\ddot{\theta}}\beta^2}{\beta_V^2} = 1 \to \beta_{\ddot{\theta}} = \beta_V^2 / \beta^2, \tag{16}$$

$$\frac{(\Pi_9)_m}{(\Pi_9)_p} = \frac{\beta_E}{\beta_\rho \beta^3 \beta_V^2} = 1 \to \beta_E = \beta_\rho \beta^3 \beta_V^2, \tag{17}$$

$$\frac{(\Pi_{10})_m}{(\Pi_{10})_p} = \frac{\beta_I}{\beta_\rho \beta^3 \beta_V} = 1 \to \beta_I = \beta_\rho \beta^3 \beta_V, \tag{18}$$

$$\frac{(\Pi_{11})_m}{(\Pi_{11})_p} = \frac{\beta_M}{\beta_\rho \beta^3} = 1 \to \beta_M = \beta_\rho \beta^3, \tag{19}$$

$$\frac{(\Pi_{12})_m}{(\Pi_{12})_p} = \frac{\beta_{H'}}{\beta} = 1 \to \beta_{H'} = \beta, \tag{20}$$

$$\frac{(\Pi_{13})_m}{(\Pi_{13})_p} = \frac{\beta_\delta}{\beta} = 1 \to \beta_\delta = \beta, \tag{21}$$

$$\frac{(\Pi_{14})_m}{(\Pi_{14})_p} = \beta_\varepsilon = 1, \tag{22}$$

and

$$\frac{(\Pi_{15})_m}{(\Pi_{15})_p} = \beta_\theta = 1. \tag{23}$$

Eqs. (9) ~ (23) show that all physical quantities can be directly expressed by the three basic scaling factors of $\beta_\rho$, $\beta$ and $\beta_V$. Since the expressions of scaling factor in Eq. (9), Eq. (10), Eq. (12), Eq. (13), Eq. (14), Eq. (17), Eq. (19), Eq. (21) and Eq. (22) are completely identical with those in Table 2, the DLV system would have the same ability with the VSG to deal with the non-scaling problems for different materials and strain-rate-sensitive. However, the derivation procedure of these factors in DLV system is very simple without any intermediate steps, while the same procedure in VSG system needs to resort some extra relations. For example, the factor $\beta_{\sigma_d} = \beta_V^2$ can be obtained through the direct derivation of Eq. (9) in DLV system when considering the same material case (i.e., $\beta_\rho = 1$), while in VSG system the derivation would be performed through the dimensionless number $\frac{\delta^3 \sigma_d}{GV_0^2}$ with the procedure of $(\delta^3 \sigma_d / GV_0^2)_m / (\delta^3 \sigma_d / GV_0^2)_p = \beta_\delta^3 \beta_{\sigma_d} / \beta_G \beta_V^2 = \beta^3 \beta_{\sigma_d} / \beta^3 \beta_V^2 = 1$ which needs to resort two extra scaling relations $\beta_\delta = \beta$ and $\beta_G = \beta^3$ [2].

## 4. Verification

In this section, four simple typical beam models subjected to mass impact or impulsive velocity are chosen to verify the features of the DLV dimensionless system in directly matching the dimensionless expression of response equations and their corresponding scaling analysis.

### 4.1. Impact of a mass on a cantilever beam

The first structure we studied is a cantilever with length of $L$, width of $B$ and height of $H$, and it is struck at the free end by a concentrated mass $G$ with an initial impact velocity $V_0$, as shown in Fig.1.

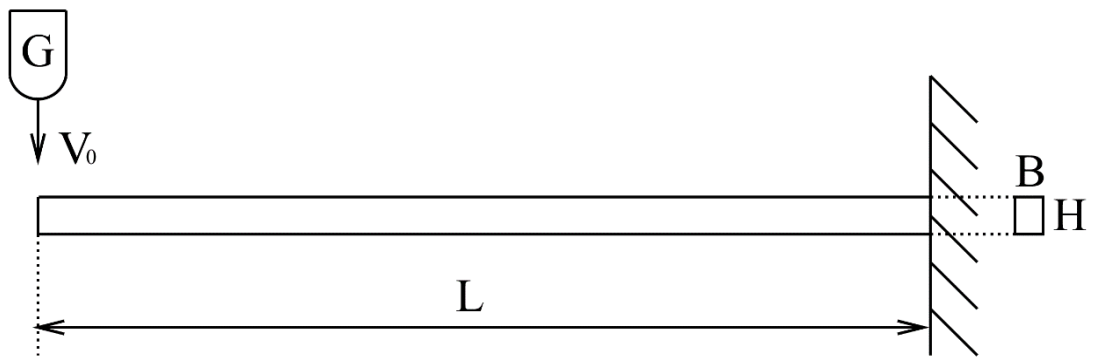

Fig. 1. A cantilever beam subject to impact mass at the free end.

### 4.1.1. Response equations

Parkes [17-18] carried out the model based theoretical research on a perfectly plastic material. The response function of the final displacement $W_f$ at the free end was given as

$$W_f = \frac{\rho' V_0^2 L^2 \gamma^2}{3\varphi_0}\left[\frac{1}{1+2\gamma}+2\ln\left(1+\frac{1}{2\gamma}\right)\right], \tag{24}$$

where $\rho' = \rho BH$ was material density per unit length, $\varphi_0 = \sigma_d BH^2/4$ was fully plastic bending moment and $\gamma = G/\rho'L$ was the mass ratio of the concentrated mass to the structure mass of the cantilever. The final rotation angle at the root was given as

$$\theta_f = \frac{\rho' L V_0^2}{6\varphi_0}(1+3\gamma)\left(1+\frac{1}{2\gamma}\right)^{-2}. \tag{25}$$

The response equation of final time was given as

$$T_f = \frac{GV_0 L}{\varphi_0} \tag{26}$$

### 4.1.2. Dimensionless expression and scaling analysis

Firstly, we rewrite the response equation of Eq. (24) to a dimensionless form,

$$\frac{W_f}{L} = \frac{4}{3}\frac{\rho V_0^2}{\sigma_d}\frac{L}{H}\gamma^2\left[\frac{1}{1+2\gamma}+2\ln\left(1+\frac{1}{2\gamma}\right)\right]. \tag{27}$$

If we regard the term of $\rho V_0^2/\sigma_d$ as $\Pi_1$, the term of $H/L$ as $\Pi_{12}$, the term of $\gamma = G/\rho LBH$ as $\Pi_{11}$ and the term of $W_f/L$ as $\Pi_{13}$, Eq. (27) reflects a functional relationship among the dimensionless numbers $\Pi_1$, $\Pi_{12}$, $\Pi_{11}$ and $\Pi_{13}$.

Secondly, we rewrite the final rotation angle equation of Eq. (25) to a dimensionless form,

$$\theta_f = \frac{2}{3}\frac{\rho V_0^2}{\sigma_d}\frac{L}{H}(1+3\gamma)\left(1+\frac{1}{2\gamma}\right)^{-2}. \qquad (28)$$

If we regard the term of $\rho V_0^2/\sigma_d$ as $\Pi_1$, the term of $H/L$ as $\Pi_{12}$, the term of $\gamma = G/\rho LBH$ as $\Pi_{11}$ and the term of $\theta_f$ as $\Pi_{15}$, Eq. (28) reflects a functional relationship among the dimensionless numbers $\Pi_1$, $\Pi_{12}$, $\Pi_{11}$ and $\Pi_{15}$.

Thirdly, we rewrite the final time equation of Eq. (26) to a dimensionless form,

$$\frac{T_f V_0}{L} = 4\frac{\rho V_0^2}{\sigma_d}\frac{L}{H}\gamma. \qquad (29)$$

If we regard the term of $\rho V_0^2/\sigma_d$ as $\Pi_1$, the term of $H/L$ as $\Pi_{12}$, the term of $\gamma = G/\rho LBH$ as $\Pi_{11}$ and the term of $T_f V_0/L$ as $\Pi_4$, Eq. (29) reflects a functional relationship among the dimensionless numbers $\Pi_1$, $\Pi_{12}$, $\Pi_{11}$ and $\Pi_4$.

The above analysis shows that the numbers $\Pi_1 = \rho V_0^2/\sigma_d$, $\Pi_{12} = H/L$ and $\Pi_{11} = G/\rho LBH$ govern the final deformation response $\Pi_{13} = W_f/L$, the rotation angle response $\Pi_{15} = \theta_f$ and the final time response $\Pi_4 = T_f V_0/L$.

When we conduct a scaling testing for this cantilever, the input parameters consisting three aspects of the material properties (including material density $\rho$ and dynamic flow stress $\sigma_d$), the geometry size (including the length $L$, the width $B$ and the height $H$) and the external loads (impact mass $G$ with its initial velocity $V_0$) could be considered. It is obvious that when the scaling relations of these input parameters satisfy $\beta = \beta_L$, $\beta_B = \beta$, $\beta_H = \beta$, $\beta_G = \beta_\rho \beta^3$ and $\beta_V = \sqrt{\beta_{\sigma_d}/\beta_\rho}$, the final responses of the beam should be completely scaled by the relations of $\beta_{W_f} = \beta$, $\beta_{\theta_f} = 1$ and $\beta_T = \beta/\beta_V$.

From the equation analysis of the cantilever model, the DLV dimensionless number system shows its property to directly match dimensionless response equation of Eqs. (27), (28) and (29). However, in VSG system the term of $W_f$ in Eq. (24) would be rewrite to the form of $W_f/(GV_0^2/\sigma_d)^{1/3}$, which is a little complicated and difficult to be understood. The main reason for the different rewritten forms is the lacking of essential physical quantities of density and length in the VSG base.

4.2. Simply supported beam subjected to impulsive loading

The second structure to be verified is a simply supported rectangular beam with length of $2L$, height of $H$ and an initial impulsive velocity $\bar{V}_0$ on the entire span, as shown in Fig.2.

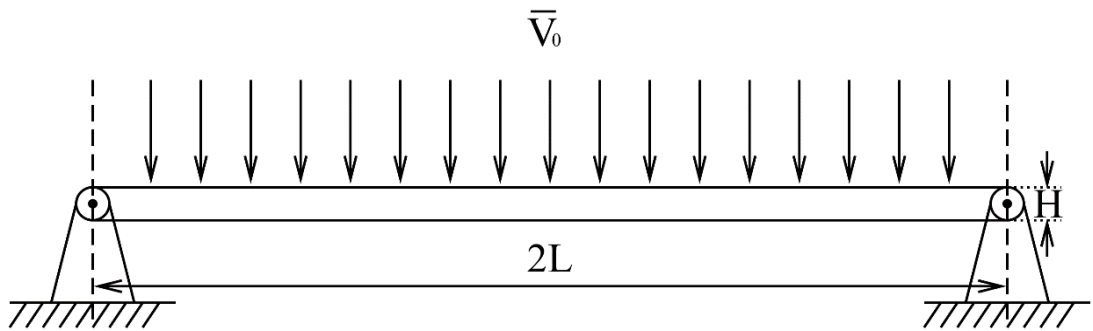

Fig. 2. A simply supported beam subject to impulsive velocity.

4.2.1. Response equations

Zhao [19] carried out the model based theoretical research on a perfectly plastic material and taking into account the influence of finite displacements. The response function of the final maximum dimensionless displacement $W_f/H$ at the mid-span was given as

$$\frac{W_f}{H} = \frac{1}{4}\left\{\left[\frac{1}{2} + 8\frac{\rho \bar{V}_0^2 L^2}{\sigma_d H^2} + \frac{1}{2}\left(1 + \frac{32}{3}\frac{\rho \bar{V}_0^2 L^2}{\sigma_d H^2}\right)^{1/2}\right]^{1/2} - 1\right\}, \tag{30}$$

And the response equation of time was given as

$$T = \frac{H}{8\bar{V}_0}\left[\left(1 + \frac{32}{3}\frac{\rho \bar{V}_0^2 L^2}{\sigma_d H^2}\right)^{1/2} - 1\right]. \tag{31}$$

4.2.2. Dimensionless expression and scaling analysis

Firstly, we rewrite the response equation of Eq. (30) to a dimensionless form

$$\frac{W_f}{L} = \frac{1}{4}\frac{H}{L}\left\{\left[\frac{1}{2} + 8\frac{\rho \bar{V}_0^2}{\sigma_d}\left(\frac{L}{H}\right)^2 + \frac{1}{2}\left(1 + \frac{32}{3}\frac{\rho \bar{V}_0^2}{\sigma_d}\left(\frac{L}{H}\right)^2\right)^{1/2}\right]^{1/2} - 1\right\}. \tag{32}$$

Similar to the Eq. (27), if we regard the term of $\rho \bar{V}_0^2/\sigma_d$ as $\Pi_1$, the term of $H/L$ as $\Pi_{12}$ and the term of $W_f/L$ as $\Pi_{13}$, Eq. (32) reflects a functional relationship among the dimensionless numbers $\Pi_1$, $\Pi_{12}$ and $\Pi_{13}$.

Secondly, we rewrite the time equation of Eq. (31) to a dimensionless form

$$\frac{T\bar{V}_0}{L} = \frac{1}{8}\frac{H}{L}\left\{\left[1 + \frac{32}{3}\frac{\rho \bar{V}_0^2}{\sigma_d}\left(\frac{L}{H}\right)^2\right]^{1/2} - 1\right\}. \tag{33}$$

Similar to the Eq. (29), If we regard the term of $\rho \bar{V}_0^2/\sigma_d$ as $\Pi_1$, the term of $H/L$ as $\Pi_{12}$ and the term of $T\bar{V}_0/L$ as $\Pi_4$, Eq. (33) reflects a functional relationship between the dimensionless numbers $\Pi_1$, $\Pi_{12}$ and $\Pi_4$.

When we conduct a scaling testing for this simply supported beam with input parameters of three aspects as the first cantilever model, the final responses should be completely scaled by the relations of $\beta_{W_f} = \beta$ and $\beta_T = \beta/\beta_V$ if the relations of input parameters satisfy $\beta = \beta_L$, $\beta_H = \beta$ and $\beta_V = \sqrt{\beta_{\sigma_d}/\beta_\rho}$. The differences between these two models are the input parameters and its scaling relations to the impact mass $G$.

## 4.3. Clamped beam subject to impulsive velocity

The third structure to be verified is a rectangular cross-section clamped beam with length of $2L$, height of $H$ and an initial impulsive velocity $\bar{V}_0$ on the entire span, as shown in Fig.3.

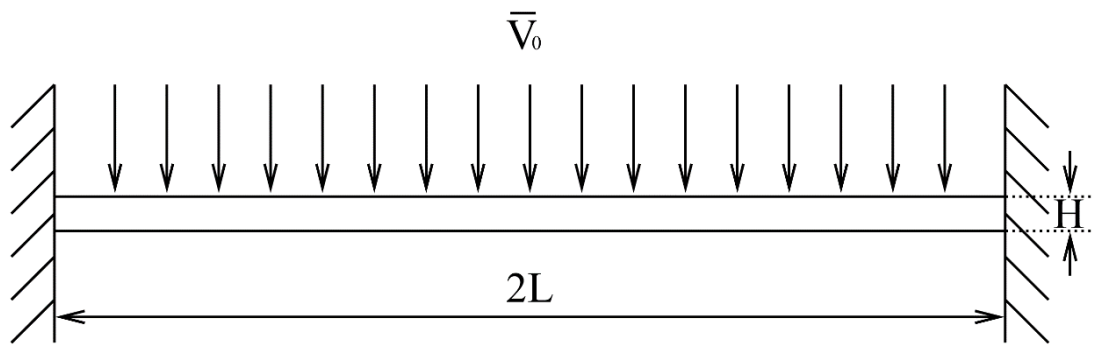

Fig. 3. A clamped beam subject to impulsive velocity.

### 4.3.1. Response equations

Jones [1] carried out the model based theoretical research on a perfectly plastic material and taking into account the influence of finite displacements. The response function of the final maximum dimensionless displacement $W_f/H$ at the mid-span was given as.

$$\frac{W_f}{H} = \frac{1}{2}\left[\left(1+\frac{3\rho \bar{V}_0^2 L^2}{\sigma_d H^2}\right)^{1/2} - 1\right]. \tag{34}$$

And the average strain rate equation was given as

$$\dot{\varepsilon} = \frac{\bar{V}_0 W_f}{3\sqrt{2}L^2}. \tag{35}$$

### 4.3.2. Dimensionless expression and scaling analysis

Firstly, we rewrite the response equation of Eq. (34) to a dimensionless form,

$$\frac{W_f}{L} = \frac{1}{2}\frac{H}{L}\left[\left(1+3\frac{\rho\bar{V}_0^2}{\sigma_d}\left(\frac{L}{H}\right)^2\right)^{1/2} - 1\right]. \tag{36}$$

Similar functional relationship among the dimensionless numbers $\Pi_1$, $\Pi_{12}$ and $\Pi_{13}$ can be found for the impulsive velocity loading.

Secondly, we rewrite the strain rate equation of Eq. (35) to a dimensionless form,

$$\frac{\dot{\varepsilon}L}{\bar{V}_0} = \frac{1}{3\sqrt{2}}\frac{W_f}{L}. \tag{37}$$

And substituting Eq. (36) into Eq. (37), it becomes

$$\frac{\dot{\varepsilon}L}{\bar{V}_0} = \frac{1}{6\sqrt{2}}\frac{H}{L}\left[\left(1+3\frac{\rho\bar{V}_0^2}{\sigma_d}\left(\frac{L}{H}\right)^2\right)^{1/2} - 1\right]. \tag{38}$$

If we regard the term of $\rho\bar{V}_0^2/\sigma_d$ as $\Pi_1$, the term of $H/L$ as $\Pi_{12}$ and the term of $\dot{\varepsilon}L/\bar{V}_0$ as $\Pi_5$, Eq. (38) reflects a functional relationship among the dimensionless numbers $\Pi_1$, $\Pi_{12}$ and $\Pi_5$. The strain rate related dimensionless number $\Pi_5$ appears.

The above analysis shows that the numbers $\Pi_1 = \rho\bar{V}_0^2/\sigma_d$ and $\Pi_{12} = H/L$ govern the final deformation response $\Pi_{13} = W_f/L$ and the final strain-rate response $\Pi_5 = \dot{\varepsilon}L/\bar{V}_0$. When we conduct a scaling testing for this clamped beam, if the relations of input parameters satisfy $\beta = \beta_L$, $\beta_H = \beta$ and $\beta_V = \sqrt{\beta_{\sigma_d}/\beta_\rho}$, the final responses should be completely scaled by the relations of $\beta_{W_f} = \beta$ and $\beta_{\dot{\varepsilon}} = \beta_V/\beta$. It is also good to find that the input and output relations of this clamped beam for the strain-rate-sensitive material was verified by the numerical calculation in Refs. [2] and [4].

4.4. Clamped beam struck at mid span

The fourth structure we studied is a rectangular cross-section beam with length of $2L$, width of $B$ and height of $H$, and it is struck at mid span by a concentrated mass $G$ with an initial impact velocity $V_0$, as shown in Fig. 4.

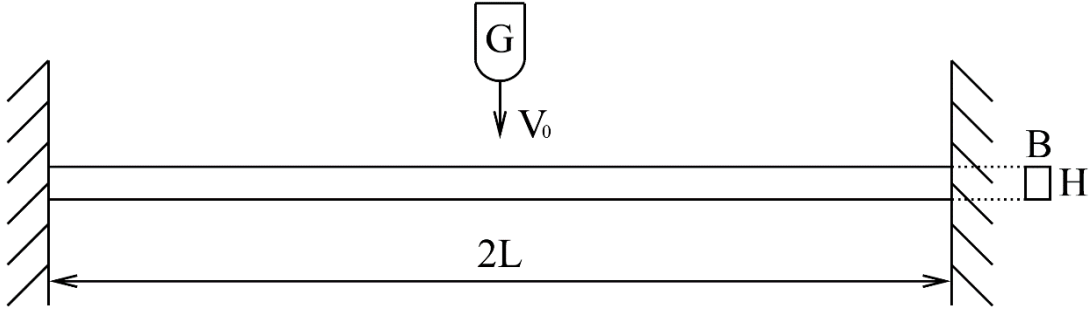

Fig. 4. A clamped beam subject to impact mass.

4.4.1. Response equations

Liu and Jones [20] carried out the beam struck at any point of the span on a perfectly plastic material and taking into account the influence of the finite displacements. For the case of a large mass $G$ (relative to small mass of the beam) at mid span, the final maximum dimensionless displacement $W_f/H$ at the loading point was expressed as,

$$\frac{W_f}{H} = \frac{1}{2}\left[\left(1 + \frac{2GV_0^2 L}{BH^3 \sigma_d}\right)^{1/2} - 1\right]. \tag{39}$$

As for true equivalent strain, the expression at any point on the span was derived by Alves and Jones [21]. And it can be expressed at the mid span as follows,

$$\varepsilon_{eq} = \begin{cases} \sqrt{\left(\dfrac{6W_H}{\gamma^2}\right)^2 + \left(\dfrac{4k}{\sqrt{3}}\dfrac{W_H}{\gamma}\right)^2}, & \text{for } W_f/H \leq 1 \\ \sqrt{\left[\left(\dfrac{W_H}{\gamma}\right)^2 + \dfrac{5}{\gamma^2}\right]^2 + \left(\dfrac{4k}{\sqrt{3}}\dfrac{W_H}{\gamma}\right)^2}, & \text{for } W_f/H > 1 \end{cases} \quad (40)$$

where $\gamma = L/H$, $W_H = W_f/H$ and $k'$ is a dimensionless constant that accounts the influence of transverse shear and is assumed to be same for different materials.

The expression of the equivalent strain rate was adopted in Refs.[4] and [21], and it can be expressed at mid span as follows,

$$\dot{\varepsilon}_{eq} = \begin{cases} \dfrac{V_0}{L}\left[\dfrac{9}{2}\left(\dfrac{H}{L}\right)^2 + \dfrac{8k^2}{3}\right]^{1/2}, & \text{for } W_f/H \leq 1 \\ \dfrac{V_0}{L}\left[\dfrac{1}{2}\left(\dfrac{W_f}{L}\right)^2 + \dfrac{8k^2}{3}\right]^{1/2}, & \text{for } W_f/H > 1 \end{cases} \quad (41)$$

4.4.2. Dimensionless expression and scaling analysis

It can be learned from the above three models, for the concentrated mass impact loading, the beam response equations of deformation, strain and strain rate would reflect functional relations between numbers $\Pi_1$, $\Pi_{11}$, $\Pi_{12}$ and $\Pi_{13}$, $\Pi_{14}$ and $\Pi_5$ respectively, which could be easily verified by same equation analysis for Eqs. (39) ~ (41). For the scaling relations, if input parameters satisfies $\beta = \beta_L$, $\beta_B = \beta$, $\beta_H = \beta$, $\beta_G = \beta_\rho \beta^3$ and $\beta_V = \sqrt{\beta_{\sigma_d}/\beta_\rho}$, the final responses of the beam should be completely scaled with the relation of $\beta_{W_f} = \beta$, $\beta_\varepsilon = 1$ and $\beta_{\dot{\varepsilon}} = \beta_V/\beta$. It is also good to find that the input and output relations of this beam for considering material strain-rate-sensitive has been verified by numerical calculation in Ref. [4]. This model in addition proves the applicable relations for simultaneously considering the effect of material strain hardening.

From the above four examples, the structure dynamic responses of deformation, angle, time, strain and strain rate of different impact problems all could be presented directly to dimensionless expression by the newly proposed dimensionless number system for different loadings of concentrated mass impact and impulsive velocity. At the same time, when we conducting a scaling testing, the scaling relations of these input and output parameters can be simultaneously obtained by writing these dimensionless numbers to the form of scaling factors through equation analysis.

## 5. Conclusions

A new group of dimensionless numbers termed DLV system is suggested in this paper in order to represent the scaled behavior of impacted structures for rigid-plastic materials with strain-hardening effects and strain-rate-sensitive. It is obtained by means of Buckingham $\Pi$ theorem with Density-Length-Velocity as the essential basis and is verified by four classical impacted model with equation analysis. Compared with the previous dimensionless system, the results have been shown the advancements of this group of numbers in clear physical significance, naturally including the well-known damage number and response number, the directly matching dimensionless expression of response equations, the ability for the addressing non-scalability etc. At the same time, the numbers in DLV system have been proved to be very important in dimensionless expression and scaling analysis for the structure dynamic responses. Because of its simple, clear and efficient properties, this newly proposed DLV (Density-Length-Velocity) dimensionless number system is suggested to be an alternative system to represent the scaled behavior of structures under impact loads.

**Acknowledgements**

The authors would like to acknowledge Mr. Xiaochuan LIU, Mr. Xulong XI and Mr. Jijun Liu (Aviation Key Laboratory of Science and Technology on Structures Impact Dynamics，Aircraft Strength Research Institute of China，Xi'an 710065，China) for their writing assistance. This research did not receive any specific grant from funding agencies in the public, commercial, or not-for-profit sectors.